\documentclass[12pt]{article}
\usepackage{enumerate}
\usepackage{amsfonts}
\usepackage{amssymb}

\newcounter{defin}  \newcounter{lemma}  \newcounter{theorem}
\newcounter{property} \newcounter{corol}  \newcounter{remark} \newcounter{example}

\newenvironment{lemma}{\par\refstepcounter{lemma}
     \textbf{Lemma \thelemma.} }{\rm\par}

\newenvironment{property}{\par\refstepcounter{property}
     \textbf{Proposition \theproperty.}\ }{\rm\par}
\newenvironment{corollary}{\par\refstepcounter{corol}
     \textbf{Corollary \thecorol.} }{\rm\par}

\newenvironment{remark}{\par\refstepcounter{remark}
     \textbf{Remark \theremark.}}{\rm\par}
\newenvironment{example}{\par\refstepcounter{example}
     \textbf{Example \theexample.}}{\rm\par}

\begin{document}
\title{On approximation of quantum channels}
\author{M.E.Shirokov, A.S.Holevo\thanks{Work is partially supported by the program "Modern problems of theoretical
mathematics" of Russian Academy of Sciences, by RFBR grant
06-01-00164-a and by grant NSH 4129.2006.1.}\\Steklov Mathematical
Institute, Moscow, Russia}\date{} \maketitle

\section{Introduction}
Although a major attention in quantum information theory so far
was paid to finite-dimensional systems and channels, there is an
increasing interest in infinite-dimensional generalizations (see
\cite{ESP}, \cite{H-Sh-2}, \cite{H-Sh-W}, \cite{Sh-2}-\cite{Sh-7}
and references therein). In the present paper we develop an
approximation approach to infinite dimensional quantum channels
based on detailed investigation of the continuity properties of
entropic characteristics of quantum channels, related to the
classical capacity, as functions of a pair ``channel, input
state''. It appears that often it is convenient to approximate a
channel by trace-nonincreasing completely positive (CP) maps --
\textit{operations}, rather than by channels. Thus it is necessary
to generalize the definitions of the channel characteristics to
operations and to consider continuity properties of these
characteristics on the extended domain.

The essential feature of infinite dimensional channels is
discontinuity and unboundedness of the main entropic characteristics
which prevents from straightforward generalization of the results
obtained in finite dimensions. A natural way to study quantum
channels with singular characteristics is to approximate them in
appropriate topology by channels (or, more generally, by operations)
with continuous characteristics, for example, by channels with
finite dimensional output space. This approach was used (implicitly)
in \cite{Sh-2} to derive the strong additivity of the Holevo
capacity ($\chi$-\textit{capacity} in what follows) for some classes
of infinite dimensional channels from the corresponding finite
dimensional results and to prove that validity of the additivity
conjecture in finite dimensions implies strong additivity of the
$\chi$-capacity for all infinite-dimensional channels.

The content of this paper is as follows. Section 2 presents basic
notions and some results of previous works used in this paper. In
section 3 we consider the topology of strong convergence on the set
of all quantum operations, which appears to be a proper topology for
the purposes of approximation. It is shown that it is this topology
in which the set of all quantum operations is isomorphic to a
particular subset of states of composite system (the generalized
Choi-Jamiolkowski isomorphism). This isomorphism implies simple
compactness criterion for subsets of quantum operations. In section
4 the continuity properties of the convex closure of the output
entropy and of the $\chi$-function (the constrained $\chi$-capacity)
as functions of pair (quantum operation, input state) are explored.
Several continuity conditions are obtained. In section 5 the
obtained results are applied to the following problems:
\begin{enumerate}[1)]
  \item  continuity of the $\chi$-capacity as function of
         a channel;
  \item  strong additivity of the $\chi$-capacity for infinite
  dimensional channels;
  \item  the representation for the convex closure of the output entropy
  of arbitrary quantum channel.
\end{enumerate}

Thus approximation of infinite dimensional quantum channels by
operations in the topology of strong convergence appears as a
useful tool in study the characteristics related to the classical
capacity. In subsequent work we plan to apply it to other
characteristics of quantum channels, such as entanglement-assisted
capacity and quantum capacity.

\section{Preliminaries}

Let $\mathcal{H}$ be a separable Hilbert space,
$\mathfrak{B}(\mathcal{H})$ -- the set of all bounded operators on
$\mathcal{H}$, $\mathfrak{T}( \mathcal{H})$ -- the Banach space of
all trace-class operators with the trace norm $\Vert\cdot\Vert_{1}$.
Let
$$
\mathfrak{T}_{1}(\mathcal{H})=\{A\in\mathfrak{T}(\mathcal{H})\,|\,A\geq
0,\mathrm{Tr}A\leq
1\}\;\;\textup{and}\;\;\mathfrak{S}(\mathcal{H})=\{A\in\mathfrak{T}_{1}(\mathcal{H})\,|\,\mathrm{Tr}A=1\}
$$
be the closed convex subsets of $\mathfrak{T}(\mathcal{H})$, which
are complete separable metric spaces with the metric defined by
the trace norm. Operators in $\mathfrak{S}(\mathcal{H})$ are
called density operators. Each density operator uniquely defines a
normal state on $\mathfrak{B}(\mathcal{H})$ \cite{B&R}, so, in
what follows we will also for brevity use the term "state".

We denote by $\mathrm{co}\mathcal{A}$ ($\overline{\mathrm{co}}
\mathcal{A}$) the convex hull (closure) of a set $\mathcal{A}$ and
by $\mathrm{co}f$ ($\overline{\mathrm{co}}f$) the convex hull
(closure) of a function $f$ \cite{J&T}. We denote by
$\mathrm{extr}\mathcal{A}$ the set of all extreme points of a
convex set $\mathcal{A}$.

Let $\mathcal{P}(\mathcal{A})$ be the set of all Borel probability
measures on complete separable metric space $\mathcal{A}$ endowed
with the topology of weak convergence \cite{Par}. This set can be
considered as a complete separable metric space as well
\cite{Par}. The subset of $\mathcal{P}(\mathcal{A})$ consisting of
measures with finite support will be denoted by
$\mathcal{P}^{\mathrm{f}}(\mathcal{A})$. In what follows we will
also use the abbreviations
$\mathcal{P}=\mathcal{P}(\mathfrak{S}(\mathcal{H}))$ and
$\widehat{\mathcal{P}}=\mathcal{P}(\mathrm{extr}\mathfrak{S}(\mathcal{H}))$.

The \textit{barycenter} of the measure $\mu\in\mathcal{P}$ is the
state defined by the Bochner integral
\[
\bar{\rho}(\mu)=\int_{\mathfrak{S}(\mathcal{H})}\sigma
\mu(d\sigma).
\]

For arbitrary subset $\mathcal{A}\subset\mathfrak{S}(\mathcal{H})$
let $\mathcal{P}_{\mathcal{A}}$ (corresp.
$\widehat{\mathcal{P}}_{\mathcal{A}}$) be the subset of
$\mathcal{P}$ (corresp. $\widehat{\mathcal{P}}$) consisting of all
measures with the barycenter in $\mathcal{A}$.

A collection of states $\{\rho_{i}\}$ with corresponding
probability distribution $\{\pi_{i}\}$ is conventionally called
\textit{ensemble} and is denoted by $\{\pi _{i},\rho _{i}\}$.  In
this paper we will consider ensemble of states as a partial case
of probability measure, so that notation
$\{\pi_{i},\rho_{i}\}\in\mathcal{P}_{\{\rho\}}$ means that
$\rho=\sum_{i}\pi_{i}\rho_{i}$.

We will use the following two extensions of the von Neumann
entropy $S(\rho)=-\mathrm{Tr}\rho\log\rho$ of a state $\rho$ to
the set $\mathfrak{T}_{1}(\mathcal{H})$ (cf.\cite{L-2})
$$
S(A)=-\mathrm{Tr}A\log A\quad\textup{and}\quad
H(A)=S(A)-\eta(\mathrm{Tr}A),\quad \forall A
\in\mathfrak{T}_{1}(\mathcal{H}),
$$
where $\eta(x)=-x\log x$.

Nonnegativity, concavity and lower semicontinuity of the von
Neumann entropy $S$ on the set $\mathfrak{S}(\mathcal{H})$ imply
the same properties of the functions $S$ and $H$ on the set
$\mathfrak{T}_{1}(\mathcal{H})$. We will use the following
properties
\begin{eqnarray}
H(\lambda A)=\lambda H(A),\quad
A\in\mathfrak{T}_{1}(\mathcal{H}),\;\lambda\geq
0,\quad\quad\quad\quad\quad\quad\quad \label{H-fun-eq}\\
H(A)+H(B-A)\leq H(B)\leq H(A)+H(B-A)+\mathrm{Tr}B
h_{2}\left(\frac{\mathrm{Tr}A}{\mathrm{Tr}B}\right)\label{H-fun-ineq},
\end{eqnarray}
where $A,B\in\mathfrak{T}_{1}(\mathcal{H}),\; A\leq B,$ and
$h_{2}(x)=\eta(x)+\eta(1-x)$.

Subadditivity property of the quantum entropy implies the
following inequality
\begin{equation}\label{gen-subadd}
S(C)\leq
S(\mathrm{Tr}_{\mathcal{H}}C)+S(\mathrm{Tr}_{\mathcal{K}}C)-\eta({\mathrm{Tr}C}),\quad\forall
C\in \mathfrak{T}_{1}(\mathcal{H}\otimes\mathcal{K}).
\end{equation}

The relative entropy for two operators $A$ and $B$ in
$\mathfrak{T}_{1}(\mathcal{H})$ is defined by (cf.\cite{L-2})
$$
H(A\,\|B)=\sum_{i}\langle i|\,(A\log A-A\log
B+B-A)\,|i\rangle
$$
where $\{|i\rangle\}$ is the orthonormal basis of eigenvectors of
$A$.

Let $\mathcal{H},\mathcal{H}^{\prime }$ be a pair of separable
Hilbert spaces which we call correspondingly input and output space.
A quantum operation $\Phi$ is a linear positive trace-nonicreasing
map from $\mathfrak{T}(\mathcal{H})$ to
$\mathfrak{T}(\mathcal{H}^{\prime })$ such that the dual map
$\Phi^{\ast}:\mathfrak{B}(\mathcal{H}^{\prime})\mapsto\mathfrak{B}(\mathcal{H})$
is completely positive. The convex set of all quantum operations
from $\mathfrak{T}(\mathcal{H})$ to
$\mathfrak{T}(\mathcal{H}^{\prime })$ will be denoted by
$\mathfrak{F}_{\leq 1}(\mathcal{H},\mathcal{H}^{\prime})$. If $\Phi$
is trace preserving then it is called quantum channel. The convex
set of all channels from $\mathfrak{T}(\mathcal{H})$ to
$\mathfrak{T}(\mathcal{H}^{\prime})$ will be denoted by
$\mathfrak{F}_{=1}(\mathcal{H},\mathcal{H}^{\prime})$.

Since the functions  $\rho\mapsto H_{\Phi}(\rho)=H(\Phi(\rho))$,
$\rho\mapsto S_{\Phi}(\rho)=S(\Phi(\rho))$ and $\rho\mapsto H(\Phi
(\rho )\Vert A)$, where $\Phi$ is a given quantum operation in
$\mathfrak{F}_{\leq 1}(\mathcal{H},\mathcal{H}^{\prime})$ and $A$ is
a given operator in $\mathfrak{T}_{1}(\mathcal{H})$, are nonnegative
and lower semicontinuous on the set $\mathfrak{S}(\mathcal{H})$, the
functionals
$$
\hat{H}_{\Phi}(\mu)=\int_{\mathfrak{S}(\mathcal{H})}H_{\Phi}(\rho)\mu(d\rho),\quad
\hat{S}_{\Phi}(\mu)=\int_{\mathfrak{S}(\mathcal{H})}S_{\Phi}(\rho)\mu(d\rho)\quad
$$
and
$$
\displaystyle\chi_{\Phi}(\mu)=\int_{\mathfrak{S}(\mathcal{H})}H(\Phi
(\rho )\Vert \Phi (\bar{\rho}(\mu)))\mu(d\rho )
$$
are well defined on the set $\mathcal{P}$.

\begin{property}\label{main-functionals-ls}
\textit{The functionals $\hat{H}_{\Phi}(\mu)$, $\hat{S}_{\Phi}(\mu)$
and $\chi _{\Phi }(\mu)$ are lower semicontinuous on the set
$\mathcal{P}$. If $S_{\Phi}(\bar{\rho}(\mu))<+\infty$ then}
\begin{equation}\label{formula}
\chi_{\Phi}(\mu)= S_{\Phi}(\bar{\rho}(\mu))-\hat{S}_{\Phi}(\mu).
\end{equation}
\end{property}

This proposition can be proved by obvious modification of the
arguments used in the proof of proposition 1 in \cite{H-Sh-2}.

\begin{corollary}\label{main-functionals-c}
\textit{Let $\mathcal{P}_{0}$ be such subset of $\mathcal{P}$ that
the function $S_{\Phi}$ is continuous on the set
$\{\bar{\rho}(\mu)\}_{\mu\in\mathcal{P}_{0}}$. Then the
functionals $\hat{H}_{\Phi}(\mu)$, $\hat{S}_{\Phi}(\mu)$ and $\chi
_{\Phi }(\mu)$ are continuous on the set $\mathcal{P}_{0}$.}
\end{corollary}

Corollary \ref{main-functionals-c} implies in particular
continuity of the functionals $\hat{H}_{\Phi}(\mu)$,
$\hat{S}_{\Phi}(\mu)$ and $\chi_{\Phi }(\mu)$ on the set
$\mathcal{P}_{\{\rho\}}$ if $S_{\Phi}(\rho)<+\infty$.

The important characteristics of the quantum channel $\Phi$ is the
convex closure $\overline{\mathrm{co}}H_{\Phi}$ of the output
entropy $H_{\Phi}(=S_{\Phi})$ \cite{Sh-7}. In this paper we
consider the convex closures $\overline{\mathrm{co}}H_{\Phi}$ and
$\overline{\mathrm{co}}S_{\Phi}$ of the functions $H_{\Phi}$ and
$S_{\Phi}$ correspondingly for arbitrary quantum operation $\Phi$
in $\mathfrak{F}_{\leq 1}(\mathcal{H},\mathcal{H}')$.

\begin{property}\label{prop-2}
\textit{Let $\Phi$ be an arbitrary quantum operation in
$\mathfrak{F}_{\leq 1}(\mathcal{H},\mathcal{H}')$ and $\rho$ be an
arbitrary state in $\mathfrak{S}(\mathcal{H})$.}

A) \textit{The following expressions hold
\begin{equation}\label{H-fun-rep-g}
\overline{\mathrm{co}}H_{\Phi}(\rho)=
\inf_{\mu\in\mathcal{P}_{\{\rho\}}}\hat{H}_{\Phi}(\mu)=\inf_{\mu\in\widehat{\mathcal{P}}_{\{\rho\}}}
\hat{H}_{\Phi}(\mu)
\end{equation}
and
\begin{equation}\label{S-fun-rep-g}
\overline{\mathrm{co}}S_{\Phi}(\rho)=
\inf_{\mu\in\mathcal{P}_{\{\rho\}}}\hat{S}_{\Phi}(\mu)=
\inf_{\mu\in\widehat{\mathcal{P}}_{\{\rho\}}}\hat{S}_{\Phi}(\mu)
\end{equation}
The infima in these expressions are achieved at some measures in
$\widehat{\mathcal{P}}_{\{\rho\}}$.}

B)\textit{ The following inequalities hold
$$
\overline{\mathrm{co}}H_{\Phi}(\rho)\leq\overline{\mathrm{co}}S_{\Phi}(\rho)\leq\overline{\mathrm{co}}H_{\Phi}(\rho)+
\eta(\mathrm{Tr}\Phi(\rho)).
$$}

C) \textit{If $\overline{\mathrm{co}}S_{\Phi}(\rho)<+\infty$ then
$$
\{S_{\Phi}(\rho)<+\infty\}\Leftrightarrow\{\overline{\mathrm{co}}S_{\Phi}(\rho)=\mathrm{co}S_{\Phi}(\rho)\},
$$
where $\mathrm{co}S_{\Phi}$ is the convex hull of the function
$S_{\Phi}$ defined by the expression
$$
\mathrm{co}S_{\Phi}(\rho)=\inf_{\{\pi_{i},
\rho_{i}\}\in\mathcal{P}^{\mathrm{f}}_{\{\rho\}}}\sum_{i}\pi_{i}S_{\Phi}(\rho_{i}).
$$
}
\end{property}

\textbf{Proof.} All assertions in  A follow from theorem 1 in
\cite{Sh-6}.

The inequalities in B are easily deduced from the representations
in A and concavity of the function $\eta$.

The implication $\Rightarrow$ in C follows from lemma 1 in
\cite{H-Sh-2} and corollary 1. Since the set of all states $\rho$
with finite $S_{\Phi}(\rho)$ is convex, $S_{\Phi}(\rho)=+\infty$
implies $\mathrm{co}S_{\Phi}(\rho)=+\infty$. This observation
proves the implication $\Leftarrow$ in C. $\square$

The $\chi$-function of the channel $\Phi$ is defined by the
expression (cf.\cite{H-Sh-1},\cite{H-Sh-2})
\begin{equation}\label{chi-fun-def}
\chi_{\Phi}(\rho)=\sup_{\{\pi_{i},
\rho_{i}\}\in\mathcal{P}^{\mathrm{f}}_{\{\rho\}}}\chi_{\Phi}(\{\pi_{i},
\rho_{i}\})=\sup_{\mu\in\mathcal{P}_{\{\rho\}}}\chi_{\Phi}(\mu),
\end{equation}
where the last equality follows from lower semicontinuity of the
functional $\chi_{\Phi}$ and lemma 1 in \cite{H-Sh-2}.

In this paper we will consider the $\chi$-function of arbitrary
quantum operation $\Phi$ in $\mathfrak{F}_{\leq
1}(\mathcal{H},\mathcal{H}')$. By using propositions
\ref{main-functionals-ls} and \ref{prop-2} it is easy to deduce
from (\ref{chi-fun-def}) that
\begin{equation}\label{main-rel}
\chi_{\Phi}(\rho)=S_{\Phi}(\rho)-\mathrm{co}S_{\Phi}(\rho)=S_{\Phi}(\rho)-\overline{\mathrm{co}}S_{\Phi}(\rho)
\end{equation}
for arbitrary state $\rho\in\mathfrak{S}(\mathcal{H})$ such that
$S_{\Phi}(\rho)<+\infty$.

\section{The topology of strong convergence}

The set $\mathfrak{F}_{\leq 1}(\mathcal{H},\mathcal{H}')$ of all
quantum operations from $\mathfrak{T}(\mathcal{H})$ into
$\mathfrak{T}(\mathcal{H}')$ can be endowed with different
topologies, in particular, with the topology of \textit{uniform
convergence}, defined by the metric
$$
d(\Phi,\Psi)=\sup_{\rho\in\mathfrak{S}(\mathcal{H})}\|\Phi(\rho)-\Psi(\rho)\|_{1},
$$
or with the topology defined by the norm of complete boundedness
\cite{Paulsen}.

But for realization of the idea of approximation of an arbitrary
quantum channel by a sequence of quantum operations with "smooth
characteristics" described in the Introduction it is convenient to
use the weaker topology of \textit{strong convergence} on the set
$\mathfrak{F}_{\leq 1}(\mathcal{H},\mathcal{H}')$, generated by the
strong operator topology on the set of all linear bounded operators
from the Banach space $\mathfrak{T}(\mathcal{H})$ into the Banach
space $\mathfrak{T}(\mathcal{H}')$. Strong convergence of the
sequence $\{\Phi_{n}\}\subset\mathfrak{F}_{\leq
1}(\mathcal{H},\mathcal{H}')$ to the quantum operation
$\Phi_{0}\in\mathfrak{F}_{\leq 1}(\mathcal{H},\mathcal{H}')$ means
that
$$
\lim_{n\rightarrow+\infty}\Phi_{n}(\rho)=\Phi_{0}(\rho),\quad
\forall\rho\in\mathfrak{S}(\mathcal{H}).
$$
In what follows we will consider the set $\mathfrak{F}_{\leq
1}(\mathcal{H},\mathcal{H}')$ as a topological space with the
topology of strong convergence. Separability of the set
$\mathfrak{S}(\mathcal{H})$ implies that the topology of strong
convergence on the set $\mathfrak{F}_{\leq
1}(\mathcal{H},\mathcal{H}')$ is metrisable (can be defined by
some metric).

\begin{remark}\label{s-topology}
\textup{Since the operator norm of any quantum operation in
$\mathfrak{F}_{\leq 1}(\mathcal{H},\mathcal{H}')$ is $\leq 1$, it is
easy to see that the topology of strong convergence on the set
$\mathfrak{F}_{\leq 1}(\mathcal{H},\mathcal{H}')$ coincides with the
topology of uniform convergence on compact subsets of
$\mathfrak{S}(\mathcal{H})$.}$\square$
\end{remark}

The advantage of the topology of strong convergence consists in
possibility to approximate arbitrary channel $\Phi$ in
$\mathfrak{F}_{=1}(\mathcal{H},\mathcal{H}')$ by sequence of quantum
operations with finite dimensional output space, for example, by the
sequence $\{\Phi_{n}(\cdot)=P_{n}\Phi(\cdot)P_{n}\}$, where
$\{P_{n}\}$ is an arbitrary sequence of finite rank projectors in
$\mathfrak{B}(\mathcal{H}')$ increasing to the unit operator
$I_{\mathcal{H}'}$.

The following proposition shows that it is the topology of strong
convergence that makes the set of all operations to be
topologically isomorphic to the special subset of states of
composite system (generalized Choi-Jamiolkowski isomorphism
\cite{Choi}).

For given full rank state
$\sigma=\sum_{i}\lambda_{i}|i\rangle\langle i|$ in
$\mathfrak{S}(\mathcal{K})$ let $\mathfrak{T}(\sigma)$ be the
subset of $\mathfrak{T}_{1}(\mathcal{K})$ consisting of all
operators $A$ such that $\left\|\frac{\langle
i|A|j\rangle}{\sqrt{\lambda_{i}\lambda_{j}}}\right\| \leq E$,
where $E$ is the unit matrix (this means that
$\sum_{i,j}\frac{\langle
i|A|j\rangle}{\sqrt{\lambda_{i}\lambda_{j}}}|i\rangle\langle
j|\leq I_{\mathcal{K}}$).

\begin{property}\label{Y-isomorphism}
\textit{Let $\mathcal{H}$, $\mathcal{H}'$ and $\mathcal{K}$ be
separable  Hilbert spaces and $|\Omega\rangle$ be an unit vector
in $\mathcal{H}\otimes\mathcal{K}$ such that
$\sigma=\mathrm{Tr}_{\mathcal{H}}|\Omega\rangle\langle\Omega|$ is
a full rank state in $\mathcal{K}$. Then the map
$$
\mathfrak{Y}:\Phi\mapsto A_{\Phi}=\Phi\otimes
\mathrm{Id}(|\Omega\rangle\langle \Omega|)
$$
is a topological isomorphism from $\mathfrak{F}_{\leq
1}(\mathcal{H},\mathcal{H}')$ onto the subset
$$
\mathfrak{T}_{1}(\mathcal{H}')\otimes\mathfrak{T}(\sigma)=
\{A\in\mathfrak{T}_{1}(\mathcal{H}'\otimes\mathcal{K})\,|\,\mathrm{Tr}_{\mathcal{H}'}A\in
\mathfrak{T}(\sigma)\}.
$$}

\textit{The restriction of the map $\mathfrak{Y}$ to the set
$\mathfrak{F}_{=1}(\mathcal{H},\mathcal{H}')$ of channels is a
topological isomorphism from
$\mathfrak{F}_{=1}(\mathcal{H},\mathcal{H}')$ onto the subset
$$
\mathfrak{S}(\mathcal{H}')\otimes\{\sigma\}=
\{\omega\in\mathfrak{S}(\mathcal{H}'\otimes\mathcal{K})\,|\,\mathrm{Tr}_{\mathcal{H}'}\omega=\sigma\}.
$$}
\end{property}

\textbf{Proof.} The second assertion of the proposition obviously
follows from the first.

Let $\sigma=\sum_{i}\lambda_{i}|i\rangle\langle i|$ and
$|\Omega\rangle=\sum_{i}\sqrt{\lambda_{i}}|i\rangle\otimes|i\rangle$,
where $\{|i\rangle\}$ is an orthonormal basis in
$\mathcal{H}\cong\mathcal{H}'\cong\mathcal{K}$.

Let $\Phi(\cdot)=\sum_{k}V_{k}(\cdot)V_{k}^{*}$ be a quantum
operation in $\mathfrak{F}_{\leq 1}(\mathcal{H},\mathcal{H}')$ so
that $\sum_{k}V_{k}^{*}V_{k}\leq I_{\mathcal{H}}$. We have
$$
\langle i|\mathrm{Tr}_{\mathcal{H}'}\Phi\otimes
\mathrm{Id}(|\Omega\rangle\langle
\Omega|)|j\rangle=\sqrt{\lambda_{i}\lambda_{j}}\mathrm{Tr}\Phi(|i\rangle\langle
j|)=\sqrt{\lambda_{i}\lambda_{j}}\langle
i|\sum_{k}V_{k}^{*}V_{k}|j\rangle.
$$
This implies $\mathrm{Tr}_{\mathcal{H}'}\Phi\otimes
\mathrm{Id}(|\Omega\rangle\langle
\Omega|)\in\mathfrak{T}(\sigma)$.

It is clear that the map $\mathfrak{Y}$ is continuous. It is
injective since
\begin{equation}\label{aqc-3}
\Phi\otimes \mathrm{Id}(|\Omega\rangle\langle
\Omega|)=\sum_{i,j}\sqrt{\lambda_{i}\lambda_{j}}\Phi(|i\rangle\langle
j|)\otimes|i\rangle\langle j|,
\end{equation}
and hence the operator $\Phi\otimes
\mathrm{Id}(|\Omega\rangle\langle \Omega|)$ determines action of the
quantum operation $\Phi$ on the operators $|i\rangle\langle j|$ for
all $i$ and $j$. By generalizing the arguments in \cite{Horodecki}
to the infinite dimensional case we will show that for each operator
$A$ in $\mathfrak{T}_{1}(\mathcal{H}')\otimes\mathfrak{T}(\sigma)$
there exists quantum operation $\Phi_{A}$ in $\mathfrak{F}_{\leq
1}(\mathcal{H},\mathcal{H}')$ such that $A=\mathfrak{Y}(\Phi_{A})$.

Let $A=\sum_{k}\pi_{k}|\psi_{k}\rangle\langle \psi_{k}|$, where
$|\psi_{k}\rangle=\sum_{i,j}c_{ij}^{k}|i\rangle\otimes|j\rangle$
is a unit vector in $\mathcal{H}'\otimes\mathcal{K}$ for each $k$.
Let $\mathrm{Tr}_{\mathcal{H}'}A=\sum_{i,j}a_{ij}|i\rangle\langle
j|$. The equality
$$
\sum_{i,j}a_{ij}|i\rangle\langle
j|=\mathrm{Tr}_{\mathcal{H}'}A=\mathrm{Tr}_{\mathcal{H}'}\sum_{k,i,j,p,t}\!\pi_{k}
c_{ij}^{k}\overline{c_{pt}^{k}}|i\rangle\langle
p|\otimes|j\rangle\langle t|=\sum_{k,i,j,t}\pi_{k}
c_{ij}^{k}\overline{c_{it}^{k}}|j\rangle\langle t|
$$
implies that
\begin{equation}\label{aqc-1}
\sum_{k,i}\pi_{k} c_{ij}^{k}\overline{c_{it}^{k}}=a_{jt},\quad
\forall j,t,
\end{equation}
in particular
\begin{equation}\label{aqc-2}
\sum_{k,i}\pi_{k} |c_{ij}^{k}|^{2}=a_{jj},\quad \forall j.
\end{equation}

By using the condition
$\mathrm{Tr}_{\mathcal{H}'}A\in\mathfrak{T}(\sigma)$ and equality
(\ref{aqc-2}) it is easy to show that
$\pi_{k}\sum_{t}|c_{ti}^{k}|^{2}\leq\lambda_{i}$ for each $i$ and
$k$. Hence for each $k$ we can define bounded operator $V_{k}$
from $\mathcal{H}$ into $\mathcal{H}'$ by its action on the
vectors $\{|i\rangle\}$ as follows
$$
V_{k}|i\rangle=\sqrt{\frac{\pi_{k}}{\lambda_{i}}}\sum_{t}c_{ti}^{k}|t\rangle.
$$

Direct calculation shows that
$$
A=\sum_{k}V_{k}\otimes
I_{\mathcal{K}}\,|\Omega\rangle\langle\Omega|\,V_{k}^{*}\otimes
I_{\mathcal{K}}=\Phi_{A}\otimes\mathrm{Id}(|\Omega\rangle\langle
\Omega|),
$$
where $\Phi_{A}(\cdot)=\sum_{k}V_{k}(\cdot)V_{k}^{*}$ is a CP map
from $\mathfrak{T}(\mathcal{H})$ into
$\mathfrak{T}(\mathcal{H}')$.

It follows from equality (\ref{aqc-1}) that $\langle i
|\sum_{k}V_{k}^{*}V_{k}|j\rangle=\frac{a_{ij}}{\sqrt{\lambda_{i}\lambda_{j}}}$.
Hence the condition
$\mathrm{Tr}_{\mathcal{H}'}A\in\mathfrak{T}(\sigma)$ means
$\sum_{k}V_{k}^{*}V_{k}\leq I_{\mathcal{H}}$ so that
$\Phi_{A}\in\mathfrak{F}_{\leq 1}(\mathcal{H},\mathcal{H}')$.

To complete the proof it is necessary to prove openness of the map
$\mathfrak{Y}$. By using expression (\ref{aqc-3}) it is easy to
see that for any sequence $\{A_{n}\}$ of operators in
$\mathfrak{T}_{1}(\mathcal{H}')\otimes\mathfrak{T}(\sigma)$,
converging to the operator $A_{0}$, the sequence
$\{\Phi_{A_{n}}(|i\rangle\langle j|)\}$ of trace class operators
converges to the operator $\Phi_{A_{0}}(|i\rangle\langle j|)$ (in
the trace norm topology) for each $i$ and $j$. Since the operator
norm of quantum operation in $\mathfrak{F}_{\leq
1}(\mathcal{H},\mathcal{H}')$ is $\leq 1$, this implies strong
convergence of the sequence $\{\Phi_{A_{n}}\}$ to the quantum
operation $\Phi_{A_{0}}$. $\square$

\begin{remark}\label{tsc}
It follows from the proof of proposition \ref{Y-isomorphism} that
in infinite dimensions the set of \textit{all} CP maps is not
isomorphic to the set of states of composite quantum system in
contrast to the finite dimensional case
(cf.\cite{Horodecki}).$\square$
\end{remark}

Proposition \ref{Y-isomorphism} makes possible to study properties
of subsets of quantum operations (resp. channels) by identifying
these subsets with subsets of trace class operators (resp. states).
For example, it implies that the set
$\mathfrak{F}_{\sigma\mapsto\rho}$ of all channels transforming a
given full rank state $\sigma$ into a given arbitrary state $\rho$
is topologically isomorphic to the set $\mathcal{C}(\rho,\sigma)$ of
all states $\omega$ in
$\mathfrak{S}(\mathcal{H}\otimes\mathcal{H}')$ such that
$\mathrm{Tr}_{\mathcal{H}'}\omega=\sigma$ and
$\mathrm{Tr}_{\mathcal{H}}\omega=\rho$.

Proposition \ref{Y-isomorphism} provides the simple proof of the
following compactness criterion for subsets of quantum operations
in the topology of strong convergence.

\begin{corollary}\label{aqc-áomp-á}
\textup{1)} \textit{The subset
$\mathfrak{F}_{0}\subseteq\mathfrak{F}_{\leq
1}(\mathcal{H},\mathcal{H}')$ is compact if there exists full rank
state $\sigma$  in $\mathfrak{S}(\mathcal{H})$ such that
$\{\Phi(\sigma)\}_{\Phi\in\mathfrak{F}_{0}}$ is a compact subset
of $\mathfrak{T}_{1}(\mathcal{H}')$. }

\textup{2)} \textit{If the subset
$\mathfrak{F}_{0}\subseteq\mathfrak{F}_{\leq
1}(\mathcal{H},\mathcal{H}')$ is compact then the set
$\{\Phi(\sigma)\}_{\Phi\in\mathfrak{F}_{0}}$ is a compact subset
of $\mathfrak{T}_{1}(\mathcal{H}')$ for arbitrary state $\sigma$
in $\mathfrak{S}(\mathcal{H})$.}
\end{corollary}

\textbf{Proof.} 1) For arbitrary state
$\sigma=\sum_{i}\lambda_{i}|i\rangle\langle i|$ in
$\mathfrak{S}(\mathcal{K})$ the set $\mathfrak{T}(\sigma)$ is a
compact subset of $\mathfrak{T}_{1}(\mathcal{K})$. It follows from
the compactness criterion for subsets of
$\mathfrak{T}_{1}(\mathcal{K})$ (the proposition in the Appendix).
Indeed, if $P_{n}=\sum_{i=1}^{n}|i\rangle\langle i|$ then
$$
\mathrm{Tr}A(I_{\mathcal{K}}-P_{n})=\sum_{i>n}\langle
i|A|j\rangle\leq \sum_{i>n}\lambda_{i},\quad\forall
A\in\mathfrak{T}(\sigma).
$$

Hence compactness of the set $\mathfrak{F}_{0}$ in the topology of
strong convergence follows from  proposition \ref{Y-isomorphism}
and the corollary in the Appendix.

2) This assertion obviously follows from the definition of the
topology of strong convergence. $\square$

\begin{example}\label{aqc-e-1}
Let $\sigma$ be a full rank state in $\mathfrak{S}(\mathcal{H})$
and $A$ be an arbitrary operator in
$\mathfrak{T}_{1}(\mathcal{H}')$. By corollary \ref{aqc-áomp-á}
the set
$$
\mathfrak{F}_{\sigma\mapsto A}=\{\Phi\in\mathfrak{F}_{\leq
1}(\mathcal{H},\mathcal{H}')\,|\,\Phi(\sigma)=A\}
$$
is compact in the topology of strong convergence.  Note that this
set is not compact in the topology of uniform convergence. Note
also that the set of \textit{all} CP maps transforming the state
$\sigma$ into the operator $A$ is not compact in the topology of
strong convergence.
\end{example}

\begin{example}\label{aqc-e-2}
Let $\sigma$ be a full rank state in $\mathfrak{S}(\mathcal{H})$ and
$H'$ be a $\mathfrak{H}$-operator (positive operator with
eigenvalues of finite multiplicity tending to the infinity, which
can be interpreted as a Hamiltonian of a quantum system
\cite{H-Sh-2}) in the space $\mathcal{H}'$. Corollary
\ref{aqc-áomp-á} and the lemma in \cite{H-c-w-c} imply that the set
of channels
$$
\{\Phi\in\mathfrak{F}_{=1}(\mathcal{H},\mathcal{H}')\,|\,\mathrm{Tr}H'\Phi(\sigma)\leq
h\}
$$
is compact in the topology of strong convergence for each $h>0$.

Let $H$ be an arbitrary $\mathfrak{H}$-operator in the space
$\mathcal{H}$. For given $k>0$ consider the set of channels
\begin{equation}\label{sp-def}
\mathfrak{F}_{H,H',k}=\left\{\Phi\in\mathfrak{F}_{=1}(\mathcal{H},
\mathcal{H}')\,\left|\,
\sup_{\rho\in\mathfrak{S}(\mathcal{H}),\mathrm{Tr}H\rho<+\infty}
\frac{\mathrm{Tr}H'\Phi(\rho)}{\mathrm{Tr}H\rho}\leq
k\right.\right\}
\end{equation}

Considering the $\mathfrak{H}$-operators $H$ and $H'$ as the
Hamiltonians of the input and output systems correspondingly, the
set $\mathfrak{F}_{H,H',k}$ can be treated as the set of channels
with the energy amplification factor not increasing $k$. By the
above observation the set  $\mathfrak{F}_{H,H',k}$ is compact in
the topology of strong convergence for each $k$.
\end{example}

\section{Continuity properties of the entropic characteristics related to the classical capacity}

For realization of the approximation procedures described in the
Introduction it is necessary to obtain sufficient conditions for
convergence of the characteristics to be explored. In this section
we consider analytical properties of the functions $(\Phi,
\rho)\mapsto \chi_{\Phi}(\rho)$ and $(\Phi, \rho)\mapsto
\overline{\mathrm{co}}H_{\Phi}(\rho)$ defined on the Cartesian
product of the set $\mathfrak{F}_{\leq
1}(\mathcal{H},\mathcal{H}')$ of quantum operations (with the
topology of strong convergence) and the set
$\mathfrak{S}(\mathcal{H})$ (with the topology of the trace norm).

\begin{property}\label{aqc-l-s}
\textit{The functions $(\Phi, \rho)\mapsto \chi_{\Phi}(\rho)$ and
$(\Phi, \rho)\mapsto \overline{\mathrm{co}}H_{\Phi}(\rho)$ are
lower semicontinuous on the set $\mathfrak{F}_{\leq
1}(\mathcal{H},\mathcal{H}')\times\mathfrak{S}(\mathcal{H})$.}
\end{property}

\textbf{Proof.} Lower semicontinuity of the function $(\Phi,
\rho)\mapsto \chi_{\Phi}(\rho)$ can be proved by the simple
modification of the proof of lower semicontinuity of the function
$\rho\mapsto \chi_{\Phi}(\rho)$ (proposition 3 in \cite{Sh-2}).

The proof of lower semicontinuity of  the function $(\Phi,
\rho)\mapsto \overline{\mathrm{co}}H_{\Phi}(\rho)$ is based on
lemma \ref{joint-l-s} below and the compactness criterion for
subsets of $\mathcal{P}$.

Suppose that the function $(\Phi, \rho)\mapsto
\overline{\mathrm{co}}H_{\Phi}(\rho)$ is not lower semicontinuous.
This means existence of sequences
$\{\Phi_{n}\}\subset\mathfrak{F}_{\leq
1}(\mathcal{H},\mathcal{H}')$ and
$\{\rho_{n}\}\subset\mathfrak{S}(\mathcal{H})$, converging to the
operation $\Phi_{0}$ and to the state $\rho_{0}$,  such that
\begin{equation}\label{l-s-b}
\lim_{n\rightarrow+\infty}\overline{\mathrm{co}}H_{\Phi_{n}}(\rho_{n})<\overline{\mathrm{co}}H_{\Phi_{0}}(\rho_{0}).
\end{equation}

For each $n>0$ proposition \ref{prop-2} guarantees existence of
 measure $\mu_{n}\in\mathcal{P}_{\{\rho_{n}\}}$ such that
$$
\overline{\mathrm{co}}H_{\Phi_{n}}(\rho_{n})=\hat{H}_{\Phi_{n}}(\mu_{n}).
$$
By the compactness criterion for subsets of $\mathcal{P}$
(proposition 2 in \cite{H-Sh-2}) the sequence $\{\mu_{n}\}_{n>0}$
is relatively compact and hence there exists subsequence
$\{\mu_{n_{k}}\}_{k}$ converging to some measure $\mu_{0}$.
Continuity of the map $\mu\mapsto\bar{\rho}(\mu)$ implies
$\mu_{0}\in\mathcal{P}_{\{\rho_{0}\}}$. By using lemma
\ref{joint-l-s} we obtain
$$
\liminf_{k\rightarrow+\infty}\overline{\mathrm{co}}H_{\Phi_{n_{k}}}(\rho_{n_{k}})=
\liminf_{k\rightarrow+\infty}\hat{H}_{\Phi_{n_{k}}}(\mu_{n_{k}})\geq\hat{H}_{\Phi_{0}}(\mu_{0})\geq
\overline{\mathrm{co}}H_{\Phi_{0}}(\rho_{0}),
$$
which contradicts to (\ref{l-s-b}).$\square$

\begin{lemma}\label{joint-l-s}
\textit{The functional $(\Phi,\mu)\mapsto\hat{H}_{\Phi}(\mu)$ is
lower semicontinuous on the set $\mathfrak{F}_{\leq
1}(\mathcal{H},\mathcal{H}')\times\mathcal{P}$.}
\end{lemma}

\textbf{Proof.} Suppose that there exist such sequences
$\{\Phi_{n}\}\subset\mathfrak{F}_{\leq
1}(\mathcal{H},\mathcal{H}')$ and $\{\mu_{n}\}\subset\mathcal{P}$,
converging to the operation $\Phi_{0}$ and to the measure
$\mu_{0}$, that
\begin{equation}\label{l-s-b-4}
\lim_{n\rightarrow+\infty}\hat{H}_{\Phi_{n}}(\mu_{n})<\hat{H}_{\Phi_{0}}(\mu_{0}).
\end{equation}
Let $\nu_{n}=\mu_{n}\circ\Phi_{n}^{-1}$ be the image of the
measure $\mu_{n}$ under the map $\Phi_{n}$ for each $n$. By
proposition 6.1 in \cite{Par} to prove that the sequence
$\{\nu_{n}\}$ of measures in
$\mathcal{P}(\mathfrak{T}_{1}(\mathcal{H}'))$ weakly converges to
the measure $\nu_{0}=\mu_{0}\circ\Phi_{0}^{-1}$ it is sufficient
to show that
\begin{equation}\label{aqc-lim-1}
\lim_{n\rightarrow+\infty}\int_{\mathfrak{T}_{1}(\mathcal{H}')}f(A)\nu_{n}(dA)=
\int_{\mathfrak{T}_{1}(\mathcal{H}')}f(A)\nu_{0}(dA)
\end{equation}
for any bounded uniformly continuous function $f$ on the set
$\mathfrak{T}_{1}(\mathcal{H}')$.  By the construction of the
sequence $\{\nu_{n}\}$ relation (\ref{aqc-lim-1}) is equivalent to
the following one
\begin{equation}\label{aqc-lim-2}
\lim_{n\rightarrow+\infty}\int_{\mathfrak{S}(\mathcal{H})}f(\Phi_{n}(\rho))\mu_{n}(d\rho)=
\int_{\mathfrak{S}(\mathcal{H})}f(\Phi_{0}(\rho))\mu_{0}(d\rho).
\end{equation}

By Prohorov's theorem (cf. \cite{Par}) compactness of the sequence
$\{\mu_{n}\}_{n\geq 0}$ (coming with separability and completeness
of the space $\mathfrak{S}(\mathcal{H})$) implies that this sequence
is \textit{tight}, which means existence of such compact set
$\mathcal{C}_{\varepsilon}\subset\mathfrak{S}(\mathcal{H})$ for each
$\varepsilon>0$ that
$\mu_{n}(\mathcal{C}_{\varepsilon})>1-\varepsilon$ for all $n\geq
0$. For each $n$ we have
$$
\begin{array}{c}
\\\displaystyle|\int_{\mathfrak{S}(\mathcal{H})}f(\Phi_{n}(\rho))\mu_{n}(d\rho)-
\int_{\mathfrak{S}(\mathcal{H})}f(\Phi_{0}(\rho))\mu_{0}(d\rho)|\\\\\displaystyle\leq
|\int_{\mathcal{C}_{\varepsilon}}f(\Phi_{n}(\rho))\mu_{n}(d\rho)-
\int_{\mathcal{C}_{\varepsilon}}f(\Phi_{0}(\rho))\mu_{0}(d\rho)|+
2\varepsilon\sup_{A\in\mathfrak{T}_{1}(\mathcal{H})}|f(A)|\\\\\displaystyle\leq
\sup_{\rho\in\mathcal{C}_{\varepsilon}}|f(\Phi_{n}(\rho))-f(\Phi_{0}(\rho))|\\\\\displaystyle+
|\int_{\mathcal{C}_{\varepsilon}}f(\Phi_{0}(\rho))\mu_{n}(d\rho)-
\int_{\mathcal{C}_{\varepsilon}}f(\Phi_{0}(\rho))\mu_{0}(d\rho)|
+2\varepsilon\sup_{A\in\mathfrak{T}_{1}(\mathcal{H})}|f(A)|.
\end{array}
$$
The first term in the right side of this inequality tends to zero
as $n\rightarrow+\infty$ due to uniform continuity of the function
$f$ and uniform convergence of the sequence $\{\Phi_{n}\}$ to the
quantum operation $\Phi_{0}$ on the compact set
$\mathcal{C}_{\varepsilon}$, provided by strong convergence (see
remark \ref{s-topology}). The second term  tends to zero as
$n\rightarrow+\infty$ due to weak convergence of the sequence
$\{\mu_{n}\}$ to the measure $\mu_{0}$. Since $\varepsilon$ is
arbitrary this observation proves (\ref{aqc-lim-2}) and hence
(\ref{aqc-lim-1}). Weak convergence of the sequence
$\{\nu_{n}=\mu_{n}\circ\Phi_{n}^{-1}\}$ to the measure
$\nu_{0}=\mu_{0}\circ\Phi_{0}^{-1}$ and lower semicontinuity of
the functional
$\hat{H}(\nu)=\int_{\mathfrak{T}_{1}(\mathcal{H}')}H(A)\nu(dA)$ on
the set $\mathcal{P}(\mathfrak{T}_{1}(\mathcal{H}'))$ (which
follows from nonnegativity and lower semicontinuity of the
function $H(A)$ on the set $\mathfrak{T}_{1}(\mathcal{H}'$)) imply
$$
\begin{array}{c}\displaystyle
\displaystyle\liminf_{n\rightarrow+\infty}\hat{H}_{\Phi_{n}}(\mu_{n})=
\liminf_{n\rightarrow+\infty}\hat{H}(\nu_{n})\geq
\hat{H}(\nu_{0})=\hat{H}_{\Phi_{0}}(\mu_{0}),
\end{array}
$$
which contradicts to (\ref{l-s-b-4}). $\square$

By concavity of the entropy and convexity of the relative entropy
proposition \ref{aqc-l-s} implies the following observation.

\begin{corollary}\label{aqc-l-s-á}
\textit{For arbitrary state $\sigma$ in $\mathfrak{S}(\mathcal{H})$
the functions
$$
\Phi\mapsto \chi_{\Phi}(\sigma)\quad \textit{and} \quad
\Phi\mapsto \overline{\mathrm{co}}H_{\Phi}(\sigma)
$$
are lower semicontinuous convex and concave functions on the set
$\mathfrak{F}_{\leq 1}(\mathcal{H},\mathcal{H}')$
correspondingly.}
\end{corollary}

By corollary \ref{aqc-l-s-á} the function $\Phi\mapsto
\overline{\mathrm{co}}H_{\Phi}(\sigma)$ achieves its infimum on
any convex compact subset of
$\mathfrak{F}_{=1}(\mathcal{H},\mathcal{H}')$ at some extreme
point of this subset. Hence the set
$\mathfrak{F}_{\sigma\mapsto\rho}$ of all channels mapping given
full rank state $\sigma$ in $\mathfrak{S}(\mathcal{H})$ to given
state $\rho$ in $\mathfrak{S}(\mathcal{H}')$ (see example
\ref{aqc-e-1} in the previous section) contains such channel
$\Phi_{\sigma,\rho}$ that
$$
\overline{\mathrm{co}}H_{\Phi_{\sigma,\rho}}(\sigma)\leq
\overline{\mathrm{co}}H_{\Phi}(\sigma),\quad\forall \Phi \in
\mathfrak{F}_{\sigma\mapsto\rho}.
$$
If $\rho\cong\sigma$ then
$\Phi_{\sigma,\rho}(\cdot)=U(\cdot)U^{*}$ and
$\overline{\mathrm{co}}H_{\Phi_{\sigma,\rho}}(\sigma)=0$, where
$U$ is any unitary map from $\mathcal{H}$ onto $\mathcal{H}'$ such
that $U\sigma U^{*}=\rho$. In general case the channel
$\Phi_{\sigma,\rho}$ is the image of some extreme point of the
compact convex set $\mathcal{C}(\sigma,\rho)$ (defined before
corollary \ref{aqc-áomp-á}) under the map $\mathfrak{Y}^{-1}$ and
in some sense can be considered as a channel with minimal noise
transforming the state $\sigma$ into the state $\rho$.

Proposition \ref{aqc-l-s} and relation (\ref{main-rel}) imply the
following sufficient condition of continuity of the functions
$(\Phi,\rho)\mapsto\chi_{\Phi}(\rho)$ and $(\Phi,\rho)\mapsto
\overline{\mathrm{co}}H_{\Phi}(\rho)$.

\begin{property}\label{aqc-c-c}\footnote{This proposition is a generalization of proposition 7 in \cite{Sh-3}.}
\textit{Let $\{\Phi_{n}\}$ be a sequence of operations in
$\mathfrak{F}_{\leq 1}(\mathcal{H},\mathcal{H}')$ strongly
converging to the channel $\Phi_{0}$ and  $\{\rho_{n}\}$ be a
sequence of states in $\mathfrak{S}(\mathcal{H})$ converging to the
state $\rho_{0}$.} \textit{If
$$
\lim_{n\rightarrow+\infty}H_{\Phi_{n}}(\rho_{n})=H_{\Phi_{0}}(\rho_{0})<+\infty
$$
then
$$
\lim_{n\rightarrow+\infty}\!\!\overline{\mathrm{co}}H_{\Phi_{n}}(\rho_{n})=\!
\lim_{n\rightarrow+\infty}\!\!\overline{\mathrm{co}}S_{\Phi_{n}}(\rho_{n})=
\overline{\mathrm{co}}H_{\Phi_{0}}(\rho_{0})\;\; and
\lim_{n\rightarrow+\infty}\!\!\chi_{\Phi_{n}}(\rho_{n})=\chi_{\Phi_{0}}(\rho_{0}).
$$}
\end{property}

As an application of this condition consider the compact set
$\mathfrak{F}_{H,H',k}\times\mathcal{K}_{H,h}$, where
$\mathfrak{F}_{H,H',k}$ is the compact subset of
$\mathfrak{F}(\mathcal{H},\mathcal{H}')$ consisting of channels with
bounded energy amplification factor (defined in example
\ref{aqc-e-2}) and $\mathcal{K}_{H,h}$ is the compact subset of
$\mathfrak{S}(\mathcal{H})$  consisting of  states with bounded mean
energy (defined by the inequality $\mathrm{Tr}H\rho\leq h$). Assume
that $\mathrm{Tr}\exp(-\lambda H')<+\infty$ for all $\lambda>0$. By
using the observation in \cite{W} it is easy to see that the
function $(\Phi,\rho)\mapsto H_{\Phi}(\rho)$ is continuous on the
set $\mathfrak{F}_{H,H',k}\times\mathcal{K}_{H,h}$ for any $k$ and
$h$. Proposition \ref{aqc-c-c} implies that the functions
$(\Phi,\rho)\mapsto \overline{\mathrm{co}}H_{\Phi}(\rho)$ and
$(\Phi,\rho)\mapsto \chi_{\Phi}(\rho)$ are continuous on the set
$\mathfrak{F}_{H,H',k}\times\mathcal{K}_{H,h}$.

The special choice of approximating sequence makes possible to
ensure convergence of the functions
$\overline{\mathrm{co}}H_{\Phi}$, $\overline{\mathrm{co}}S_{\Phi}$
and $\chi_{\Phi}$ without reference to the output entropy.

\begin{property}\label{cont-cond}
\textit{Let $\{\Phi_{n}\}$ be a sequence of operations, strongly
converging to the channel $\Phi_{0}$. The relations}
$$
\begin{array}{c}\displaystyle
\displaystyle\lim_{n\rightarrow+\infty}\overline{\mathrm{co}}H_{\Phi_{n}}(\rho)=
\displaystyle\lim_{n\rightarrow+\infty}\overline{\mathrm{co}}S_{\Phi_{n}}(\rho)=
\overline{\mathrm{co}}H_{\Phi_{0}}(\rho)\quad and\;\,
\displaystyle\lim_{n\rightarrow+\infty}\chi_{\Phi_{n}}(\rho)=\chi_{\Phi_{0}}(\rho)
\end{array}
$$
\textit{hold for any state $\rho$ in $\mathfrak{S}(\mathcal{H})$
in the following cases:}
\begin{enumerate}[A)]
  \item \textit{$\Phi_{n}(\cdot)=P_{n}\Phi_{0}(\cdot)P_{n}$ for some sequence
  $\{P_{n}\}$ of projectors in $\mathfrak{B}(\mathcal{H}')$,
  increasing to the unit operator $I_{\mathcal{H}'}$;}
  \item \textit{$\Phi_{n}(\rho)\leq\Phi_{0}(\rho)$ for all $\rho$ in
  $\mathfrak{S}(\mathcal{H})$ (in the operator order).}
\end{enumerate}
\end{property}

\textbf{Proof.} A) For arbitrary state $\rho$ in
$\mathfrak{S}(\mathcal{H})$ lemma 3 in \cite{L-2} and monotonicity
of the relative entropy imply
$\overline{\mathrm{co}}H_{\Phi_{n}}(\rho)\leq
\overline{\mathrm{co}}H_{\Phi_{0}}(\rho)$ and
$\chi_{\Phi_{n}}(\rho)\leq \chi_{\Phi_{0}}(\rho)$ correspondingly.
Hence the limit relations in the proposition follow from proposition
\ref{aqc-l-s}.

B) For arbitrary state $\rho$ in $\mathfrak{S}(\mathcal{H})$
inequality (\ref{H-fun-ineq}) and lemma \ref{sp-ineq} below imply
$\overline{\mathrm{co}}H_{\Phi_{n}}(\rho)\leq
\overline{\mathrm{co}}H_{\Phi_{0}}(\rho)$ and
$\chi_{\Phi_{n}}(\rho)\leq
\chi_{\Phi_{0}}(\rho)+\eta(\mathrm{Tr}\Phi_{n}(\rho))+h_{2}(\mathrm{Tr}\Phi_{n}(\rho))$
correspondingly. Hence the limit relations in the proposition follow
from proposition \ref{aqc-l-s}.$\square$

\begin{lemma}\label{sp-ineq}
\textit{Let $\{\pi_{i},A_{i}\}$ and $\{\pi_{i},B_{i}\}$ be two
(finite) ensembles of operators in $\mathfrak{T}_{1}(\mathcal{H})$
such that $A_{i}\leq B_{i},\;\forall i$. Then
$$
\sum_{i}\pi_{i}H(A_{i}\,\|\,A)\leq\sum_{i}\pi_{i}H(B_{i}\,\|\,B)+\eta(\mathrm{Tr}A)+\mathrm{Tr}B
h_{2}\left(\frac{\mathrm{Tr}A}{\mathrm{Tr}B}\right),
$$
where $A=\sum_{i}\pi_{i}A_{i}$ and $B=\sum_{i}\pi_{i}B_{i}$.}
\end{lemma}

\textbf{Proof.} Suppose first that $H(B)<+\infty$. Then by using
inequality (\ref{H-fun-ineq}) and concavity of the functions $H$,
$h_{2}$ and $\eta$ we obtain
$$
\begin{array}{c}
\displaystyle\sum_{i}\pi_{i}H(B_{i}\,\|\,B)=S(B)-\sum_{i}\pi_{i}S(B_{i})=[H(B)-\sum_{i}\pi_{i}H(B_{i})]\\\displaystyle+
[\eta(\mathrm{Tr}B)-\sum_{i}\pi_{i}\eta(\mathrm{Tr}B_{i})]\geq
[H(A)-\sum_{i}\pi_{i}H(A_{i})]-
\sum_{i}\pi_{i}\mathrm{Tr}B_{i}h_{2}\left(\frac{\mathrm{Tr}A_{i}}{\mathrm{Tr}B_{i}}\right)\\\displaystyle+
[\eta(\mathrm{Tr}B)-\sum_{i}\pi_{i}\eta(\mathrm{Tr}B_{i})]+[H(B-A)-\sum_{i}\pi_{i}H(B_{i}-A_{i})]\\\displaystyle
\geq
[S(A)-\sum_{i}\pi_{i}S(A_{i})]-[\eta(\mathrm{Tr}A)-\sum_{i}\pi_{i}\eta(\mathrm{Tr}A_{i})]
-
\sum_{i}\pi_{i}\mathrm{Tr}B_{i}h_{2}\left(\frac{\mathrm{Tr}A_{i}}{\mathrm{Tr}B_{i}}\right)\\\displaystyle\geq
\sum_{i}\pi_{i}H(A_{i}\,\|\,A)-\mathrm{Tr}Bh_{2}\left(\frac{\mathrm{Tr}A}{\mathrm{Tr}B}\right)-\eta(\mathrm{Tr}A).
\end{array}
$$

In the case $H(B)=+\infty$ the above observation applied to the
ensembles $\{\pi_{i},P_{n}A_{i}P_{n}\}$ and
$\{\pi_{i},P_{n}B_{i}P_{n}\}$ for each $n$, where $\{P_{n}\}$ is
an arbitrary sequence of finite rank projectors increasing to the
unit operator $I_{\mathcal{H}}$, implies inequality
$$
\begin{array}{c}
\displaystyle\sum_{i}\pi_{i}H(P_{n}A_{i}P_{n}\,\|\,P_{n}AP_{n})\\
\displaystyle\leq\sum_{i}\pi_{i}H(P_{n}B_{i}P_{n}\,\|\,P_{n}BP_{n})+\eta(\mathrm{Tr}P_{n}A)+\mathrm{Tr}P_{n}B
h_{2}\left(\frac{\mathrm{Tr}P_{n}A}{\mathrm{Tr}P_{n}B}\right).
\end{array}
$$

By using lemma 4 in \cite{L-2} we can take the limit in this
inequality and obtain the assertion of the lemma. $\square$

\begin{remark}\label{note}   Proposition 7 in \cite{Sh-3} and
proposition \ref{cont-cond}A imply that the $\chi$-function
(corresp. the CCoOE) of arbitrary quantum channel can be represented
as the least upper bound of increasing sequence of concave (corresp.
convex) \textit{continuous} bounded  functions.
\end{remark}

\section{Applications}

\subsection{On continuity of the $\chi$-capacity as a function of
a channel}

The $\chi$-capacity of a quantum channel
$\Phi\in\mathfrak{F}_{=1}(\mathcal{H},\mathcal{H}')$ constrained
by an arbitrary subset
$\mathcal{A}\subseteq\mathfrak{S}(\mathcal{H})$ can be defined by
(cf.\cite{H-c-w-c},\cite{H-Sh-2})
\begin{equation}\label{ccap-1}
\bar{C}(\Phi,\mathcal{A})=\sup_{\{\pi_{i},\rho_{i}\}\in
\mathcal{P}^{\mathrm{f}}_{\mathcal{A}}}\sum_{i}\pi_{i}H(\Phi(\rho_{i})\|\Phi(\bar{\rho}))=\sup_{\rho\in\mathcal{A}}\chi_{\Phi}(\rho).
\end{equation}

By using lower semicontinuity of the relative entropy it is easy
to show that the function
$\mathfrak{F}_{=1}(\mathcal{H},\mathcal{H}')\ni\Phi\mapsto\bar{C}(\Phi,\mathcal{A})$
is lower semicontinuous, t.i.
\begin{equation}\label{chi-cap-ls}
\liminf_{n\rightarrow+\infty}\bar{C}(\Phi_{n},\mathcal{A})\geq
\bar{C}(\Phi_{0},\mathcal{A})
\end{equation}
for arbitrary sequence $\{\Phi_{n}\}$ of channels in
$\mathfrak{F}_{=1}(\mathcal{H},\mathcal{H}')$ strongly converging
to the channel $\Phi_{0}$. There exist examples showing that $>$
can take place in (\ref{chi-cap-ls}) even in the case of uniform
convergence of the sequence $\{\Phi_{n}\}$ to the channel
$\Phi_{0}$ and that the difference between the left and the right
sides can be arbitrary large \cite{Sh-2}.

If the sequence $\{\Phi_{n}\}$ is such that the inequality
$\bar{C}(\Phi_{n},\mathcal{A})\leq\bar{C}(\Phi_{0},\mathcal{A})$
can be proved for each $n$ then (\ref{chi-cap-ls}) implies that
\begin{equation}\label{chi-cap-c}
\lim_{n\rightarrow+\infty}\bar{C}(\Phi_{n},\mathcal{A})=
\bar{C}(\Phi_{0},\mathcal{A}).
\end{equation}
For example, by the monotonicity property of the relative entropy
this holds if $\Phi_{n}=\Pi_{n}\circ\Phi_{0}$ for each $n$, where
$\{\Pi_{n}\}$ is a sequence of channels in
$\mathfrak{F}_{=1}(\mathcal{H}',\mathcal{H}')$ strongly converging
to the noiseless channel.

The results of the previous section make possible to prove the
following continuity condition for the $\chi$-capacity.

\begin{property}\label{chi-cap-cont} \textit{Let $\{\Phi_{n}\}$ be a sequence of channels in
$\mathfrak{F}_{=1}(\mathcal{H},\mathcal{H}')$, strongly converging
to the channel $\Phi_{0}$, and  $\mathcal{A}$ be a compact subset
of $\mathfrak{S}(\mathcal{H})$.}

\textit{If
$\lim_{n\rightarrow+\infty}H_{\Phi_{n}}(\rho_{n})=H_{\Phi_{0}}(\rho_{0})<+\infty$
for arbitrary sequence $\{\rho_{n}\}$ of states in $\mathcal{A}$,
converging to the state $\rho_{0}$, then (\ref{chi-cap-c}) holds.}
\end{property}

\textbf{Proof.} To prove (\ref{chi-cap-c}) it is sufficient to
show that the assumption
$$
\lim_{n\rightarrow+\infty}\bar{C}(\Phi_{n},\mathcal{A})>\bar{C}(\Phi_{0},\mathcal{A})
$$
leads to a contradiction. For each $n$ let $\rho_{n}$ be such
state in $\mathcal{A}$ that
\begin{equation}\label{aqc-ineq}
\chi_{\Phi_{n}}(\rho_{n})>\bar{C}(\Phi_{n},\mathcal{A})-1/n.
\end{equation}
Compactness of the set $\mathcal{A}$ implies existence of
subsequence $\{\rho_{n_{k}}\}$ converging to some state
$\rho_{0}\in\mathcal{A}$. By the condition
$\lim_{k\rightarrow+\infty}H_{\Phi_{n_{k}}}(\rho_{n_{k}})=H_{\Phi_{0}}(\rho_{0})<+\infty$
and proposition \ref{aqc-c-c} implies
$$
\lim_{k\rightarrow+\infty}\chi_{\Phi_{n_{k}}}(\rho_{n_{k}})=
\chi_{\Phi_{0}}(\rho_{0})\leq\bar{C}(\Phi_{0},\mathcal{A}).
$$
This coming with (\ref{aqc-ineq}) leads to a contradiction.

By using proposition \ref{chi-cap-cont} it is possible to show
that the $\chi$-capacity of a channel with energy constraint is
continuous on the set of channels with bounded energy
amplification factor, considered in example \ref{aqc-e-2}.

\begin{corollary}\label{chi-cap-cont-2}
\textit{Let $H$ and $H'$ be $\mathfrak{H}$-operators (Hamiltonians)
in the spaces $\mathcal{H}$ and $\mathcal{H}'$ correspondingly such
that $\mathrm{Tr}\exp(-\lambda H')<+\infty$ for all $\lambda>0$. The
function $\Phi\mapsto\bar{C}(\Phi,\mathcal{K}_{H,h})$ is continuous
on the set $\mathfrak{F}_{H,H',k}$ (defined by (\ref{sp-def})).}
\end{corollary}

\textbf{Proof.} By the lemma in \cite{H-c-w-c} the set
$\mathcal{K}_{H,h}$ is compact. Let $h$ and $k$ be fixed positive
numbers. For arbitrary sequences
$\{\Phi_{n}\}\subset\mathfrak{F}_{H,H',k}$ and
$\{\rho_{n}\}\subset\mathcal{K}_{H,h}$ the sequence
$\{\Phi_{n}(\rho_{n})\}$  belongs to the set
$\mathcal{K}_{H',kh}$, on which the entropy is continuous by the
observation in \cite{W}.$\square$

For arbitrary quantum channel
$\Phi\in\mathfrak{F}_{=1}(\mathcal{H},\mathcal{H}')$ and arbitrary
\textit{convex} subset
$\mathcal{A}\subseteq\mathfrak{S}(\mathcal{H})$ such that
$\bar{C}(\Phi,\mathcal{A})<+\infty$ there exists the unique state
$\Omega(\Phi,\mathcal{A})$ in $\mathfrak{S}(\mathcal{H}')$ called
output optimal average for the $\mathcal{A}$-constrained channel
$\Phi$ (proposition 1 in \cite{Sh-2}\footnote{the case of noncompact
set $\mathcal{A}$ is considered in quant-ph/0408009.}). This state
inherits the main properties of the image of the average state of
optimal ensemble for a finite dimensional $\mathcal{A}$-constrained
channel $\Phi$ \cite{H-Sh-1}. If there exists an optimal measure
$\mu$ for the $\mathcal{A}$-constrained channel $\Phi$ (see
definition in \cite{H-Sh-2}) then
$\Omega(\Phi,\mathcal{A})=\Phi(\bar{\rho}(\mu))$. It is interesting
to note that continuity of the function
$\Phi\mapsto\bar{C}(\Phi,\mathcal{A})$ on some set of channels
implies continuity of the function
$\Phi\mapsto\Omega(\Phi,\mathcal{A})$ on this set.

\begin{property}\label{o-o-a-cont}
\textit{Let $\{\Phi_{n}\}$ be a sequence of channels in
$\mathfrak{F}_{=1}(\mathcal{H},\mathcal{H}')$, strongly converging
to the channel $\Phi_{0}$, and  $\mathcal{A}$ be a convex subset
of $\mathfrak{S}(\mathcal{H})$.}

\textit{If
$\displaystyle\lim_{n\rightarrow+\infty}\bar{C}(\Phi_{n},\mathcal{A})=\bar{C}(\Phi_{0},\mathcal{A})<+\infty$
then
$\displaystyle\lim_{n\rightarrow+\infty}\Omega(\Phi_{n},\mathcal{A})=\Omega(\Phi_{0},\mathcal{A})$.}
\end{property}

\textbf{Proof.} By proposition 1 in \cite{Sh-2} for arbitrary
$\varepsilon>0$ there exists ensemble $\{\pi_{i}, \rho_{i}\}$ with
the average state in $\mathcal{A}$ such that
\begin{equation}\label{1}
\chi_{\Phi_{0}}(\{\pi_{i}, \rho_{i}\})\geq
\bar{C}(\Phi_{0},\mathcal{A})-\varepsilon \;\;\textup{and}\;\:
\|\sum_{i}\pi_{i}\Phi_{0}(\rho_{i})-\Omega(\Phi_{0},\mathcal{A})\|_{1}<\varepsilon.
\end{equation}
Lower semicontinuity of the relative entropy implies
$$
\chi_{\Phi_{n}}(\{\pi_{i}, \rho_{i}\})\geq
\chi_{\Phi_{0}}(\{\pi_{i}, \rho_{i}\})-\varepsilon
$$
for all sufficiently large $n$. By the assumption
$$
\bar{C}(\Phi_{n},\mathcal{A})\leq\bar{C}(\Phi_{0},\mathcal{A})+\varepsilon
$$
for all sufficiently large $n$.

Thus for all sufficiently large $n$ we have
$$
0\leq \bar{C}(\Phi_{n},\mathcal{A})-\chi_{\Phi_{n}}(\{\pi_{i},
\rho_{i}\})\leq
\bar{C}(\Phi_{0},\mathcal{A})-\chi_{\Phi_{0}}(\{\pi_{i},
\rho_{i}\})+2\varepsilon\leq 3\varepsilon
$$
and by using corollary 1 in \cite{Sh-2} we obtain
\begin{equation}\label{2}
\begin{array}{c}\displaystyle
\frac{1}{2}\,\|\sum_{i}\pi_{i}\Phi_{n}(\rho_{i})-\Omega(\Phi_{n},\mathcal{A})\|_{1}^{2}\leq
H\left(\sum_{i}\pi_{i}\Phi_{n}(\rho_{i})\,\|\,\Omega(\Phi_{n},\mathcal{A})\right)\\\\\leq
\bar{C}(\Phi_{n},\mathcal{A})-\chi_{\Phi_{n}}(\{\pi_{i},
\rho_{i}\})\leq 3\varepsilon.
\end{array}
\end{equation}

By strong convergence of the sequence $\{\Phi_{n}\}$ to the
channel $\Phi_{0}$ we have
\begin{equation}\label{3}
\|\sum_{i}\pi_{i}\Phi_{n}(\rho_{i})
-\sum_{i}\pi_{i}\Phi_{0}(\rho_{i})\|_{1}\leq\varepsilon
\end{equation}
for all sufficiently large $n$.

By using (\ref{1}),(\ref{2}) and (\ref{3}) we obtain
$$
\!\begin{array}{c}\displaystyle
\|\Omega(\Phi_{n},\mathcal{A})-\Omega(\Phi_{0},\mathcal{A})\|_{1}\leq
\|\Omega(\Phi_{n},\mathcal{A})-\!\sum_{i}\pi_{i}\Phi_{n}(\rho_{i})\|_{1}\\\displaystyle\!+
\|\sum_{i}\pi_{i}\Phi_{n}(\rho_{i})
-\!\sum_{i}\pi_{i}\Phi_{0}(\rho_{i})\|_{1}+\|\sum_{i}\pi_{i}\Phi_{0}(\rho_{i})-\Omega(\Phi_{0},\mathcal{A})\|_{1}\leq
2\varepsilon+\!\sqrt{6\varepsilon}
\end{array}
$$
for all sufficiently large $n$. $\square$

\subsection{On additivity of the $\chi$-capacity}

The approximation procedure is the essential part of the proof that
additivity conjecture in finite dimensions implies strong additivity
of the $\chi$-capacity for all infinite dimensional channels
\cite{Sh-2}. It also provides possibility to derive strong
additivity of the $\chi$-capacity for two infinite dimensional
channels with one of them noiseless or entanglement-breaking from
the corresponding finite dimensional results \cite{H-Sh-1},
\cite{Shor-e-b-c}.\footnote{Note that direct generalization of the
proofs of these results to the infinite dimensional case seems to be
nontrivial. For example, the proof of theorem 2 in \cite{Shor-e-b-c}
is based on finiteness of the output entropy and on decomposition of
an arbitrary separable state into \textit{discrete} convex
combination of pure product states, which is not valid in the
infinite dimensional case \cite{H-Sh-W}.}

In \cite{Sh-7} strong additivity of the $\chi$-capacity for two
infinite dimensional channels with one of them complementary to
entanglement-breaking channel is proved \textit{under the condition}
that the output entropies of both channels are finite on the set of
pure input states. This condition seems to be essential since it is
coincidence of the output entropies of two comlementary channels on
the set of pure states that provides "transition" of the additivity
properties between pairs of complementary channels (see the proof of
theorem 1 in \cite{H-comp-ch}) and infinite values of these output
entropies prevent this transition. But the condition of finiteness
of the output entropy on the set of pure states for a given channel
is difficult to verify in general, which is a real obstacle in
application of the above result. Moreover, this condition is not
valid for large class of infinite dimensional channels. We will show
below that the approximation approach makes possible to overcome the
problem of infinite output entropies and to prove strong additivity
of the $\chi$-capacity for two infinite dimensional channels with
one of them complementary to entanglement-breaking channel even in
the case when the output entropies of these channels are everywhere
infinite.

\begin{property}\label{hf-p-4}
\textit{Let $\Phi\in\mathfrak{F}_{=1}(\mathcal{H},\mathcal{H}')$
be a channel such that its complementary channel is entanglement
breaking and $\Psi\in\mathfrak{F}_{=1}(\mathcal{K},\mathcal{K}')$
be an arbitrary channel. Then strong additivity of the
$\chi$-capacity holds for the channels $\Phi$ and $\Psi$.}
\end{property}

\textbf{Proof.} By using lemma 5 and proposition 6 in \cite{Sh-2}
it is possible to reduce the proof to the case
$\dim\mathcal{K}<+\infty$ and $\dim\mathcal{K}'<+\infty$. By
proposition 6 in \cite{Sh-2} it is sufficient to prove inequality
\begin{equation}\label{chi-fun-sub-add}
\chi_{\Phi\otimes\Psi}(\omega)\leq
\chi_{\Phi}(\omega^{\mathcal{H}})+\chi_{\Psi}(\omega^{\mathcal{K}})
\end{equation}
for arbitrary state $\omega$ in
$\mathfrak{S}(\mathcal{H}\otimes\mathcal{K})$ such that
$\mathrm{rank}\omega^{\mathcal{H}}<+\infty$. Let $\omega$ be such
a state and
$\mathcal{H}_{\omega}=\mathrm{supp}\omega^{\mathcal{H}}$ be the
corresponding finite dimensional subspace.

Let $\Phi(\rho)=\mathrm{Tr}_{\mathcal{H}''}V\rho V^{*}$, where $V$
is the Stinespring isometry from $\mathcal{H}$ into
$\mathcal{H}'\otimes\mathcal{H}''$. By the condition the
complementary channel
$\widehat{\Phi}(\rho)=\mathrm{Tr}_{\mathcal{H}'}V\rho V^{*}$ is
entanglement-breaking.

Let $\{P_{n}\}$ be an arbitrary sequence of finite rank projectors
in $\mathfrak{B}(\mathcal{H}'')$, increasing to the unit operator
$I_{\mathcal{H}''}$. Consider the quantum operations
$$
\Phi_{n}(\rho)=\mathrm{Tr}_{\mathcal{H}''}I_{\mathcal{H}'}\otimes
P_{n}\cdot V\rho V^{*}\cdot I_{\mathcal{H}'}\otimes
P_{n}=\mathrm{Tr}_{\mathcal{H}''}I_{\mathcal{H}'}\otimes
P_{n}\cdot V\rho V^{*},\quad \rho\in\mathfrak{S}(\mathcal{H}),
$$
and
$$
\widehat{\Phi}_{n}(\rho)=\mathrm{Tr}_{\mathcal{H}'}I_{\mathcal{H}'}\otimes
P_{n}\cdot V\rho V^{*}\cdot I_{\mathcal{H}'}\otimes
P_{n}=P_{n}\widehat{\Phi}(\rho)P_{n},\quad\rho\in\mathfrak{S}(\mathcal{H}).$$

Let $\widehat{\Psi}$ be the complementary channel to the channel
$\Psi$. Note that the restriction of the quantum operation
$\widehat{\Phi}_{n}$ to the set $\mathfrak{S}(\mathcal{H}_{\omega})$
is a finite dimensional entanglement-breaking
operation\footnote{This is an obvious generalization of the notion
of entanglement-breaking channel.}. By using proposition
\ref{prop-2}C and by repeating the arguments from the proof of
theorem 2 in \cite{Shor-e-b-c} it is possible to show existence of
such sequence $\{\sigma_{n}\}\subset\mathfrak{S}(\mathcal{K})$
converging to the state $\omega^{\mathcal{K}}$ that for each $n$ the
following inequality holds
\begin{equation}\label{approx-superadd-1}
\overline{\mathrm{co}}S_{\widehat{\Phi}_{n}\otimes\widehat{\Psi}}(\omega)=
\mathrm{co}S_{\widehat{\Phi}_{n}\otimes\widehat{\Psi}}(\omega)
\geq \overline{\mathrm{co}}S_{\widehat{\Phi}_{n}}
(\omega^{\mathcal{H}})+\alpha_{n}\overline{\mathrm{co}}S_{\widehat{\Psi}}
(\sigma_{n}),
\end{equation}
where
$\alpha_{n}=\inf_{\rho\in\mathfrak{S}(\mathcal{H}_{\omega})}\mathrm{Tr}\widehat{\Phi}_{n}(\rho)$.

Since
$$
S_{\widehat{\Phi}_{n}}(\rho)=S_{\Phi_{n}}(\rho),\quad \forall
\rho\in \mathrm{extr}\mathfrak{S}(\mathcal{H}),\quad
S_{\widehat{\Psi}}(\sigma)=S_{\Psi}(\sigma),\quad \forall
\sigma\in \mathrm{extr}\mathfrak{S}(\mathcal{K})
$$
and
$$
S_{\widehat{\Phi}_{n}\otimes\widehat{\Psi}}(\omega)=S_{\Phi_{n}\otimes\Psi}(\omega),\quad
\forall \omega\in
\mathrm{extr}\mathfrak{S}(\mathcal{H}\otimes\mathcal{K}),
$$
proposition \ref{prop-2}A implies that inequality
(\ref{approx-superadd-1}) is equivalent to the following one
\begin{equation}\label{approx-superadd-2}
\overline{\mathrm{co}}S_{\Phi_{n}\otimes\Psi}(\omega)\geq
\overline{\mathrm{co}}S_{\Phi_{n}}
(\omega^{\mathcal{H}})+\alpha_{n}\overline{\mathrm{co}}S_{\Psi}
(\sigma_{n}).
\end{equation}

Note that inequality (\ref{gen-subadd}) implies
\begin{equation}\label{gen-subadd-o-e}
S_{\Phi_{n}\otimes\Psi}(\omega)\leq
S_{\Phi_{n}}(\omega^{\mathcal{H}})+S(\mathrm{Tr}_{\mathcal{H}'}\Phi_{n}\otimes\Psi(\omega))-\varepsilon_{n},
\end{equation}
where
$\varepsilon_{n}=\eta(\mathrm{Tr}\Phi_{n}(\omega^{\mathcal{H}}))$.

By using (\ref{main-rel}), (\ref{approx-superadd-2}),
(\ref{gen-subadd-o-e}) and proposition \ref{prop-2}B we obtain
$$
\begin{array}{c}
\chi_{\Phi_{n}\otimes\Psi}(\omega)=S_{\Phi_{n}\otimes\Psi}(\omega)-
\overline{\mathrm{co}}S_{\Phi_{n}\otimes\Psi}(\omega)\\\\\leq
S_{\Phi_{n}}(\omega^{\mathcal{H}})-\overline{\mathrm{co}}S_{\Phi_{n}}(\omega^{\mathcal{H}})
+S(\mathrm{Tr}_{\mathcal{H}'}\Phi_{n}\otimes\Psi(\omega))-\overline{\mathrm{co}}S_{\Psi}(\sigma_{n})
\\\\+(1-\alpha_{n})\overline{\mathrm{co}}S_{\Psi}(\sigma_{n})
\leq
\chi_{\Phi_{n}}(\omega^{\mathcal{H}})+\chi_{\Psi}(\omega^{\mathcal{K}})
+[(1-\alpha_{n})S_{\Psi}(\sigma_{n})]\\\\
+[S(\mathrm{Tr}_{\mathcal{H}'}\Phi_{n}\otimes\Psi(\omega))-S_{\Psi}(\omega^{\mathcal{K}})]+
[\overline{\mathrm{co}}S_{\Psi}(\omega^{\mathcal{K}})-\overline{\mathrm{co}}S_{\Psi}(\sigma_{n})].
\end{array}
$$
The sequence of quantum operations $\{\Phi_{n}\}$ strongly converges
to the channel $\Phi$ and satisfies condition B in proposition
\ref{cont-cond}. This proposition and proposition \ref{aqc-l-s} make
possible to prove inequality (\ref{chi-fun-sub-add}) by taking the
limit in the above inequality since the terms in the square brackets
tends to zero as $n\rightarrow+\infty$ due to assumed finite
dimensionality of the spaces $\mathcal{H}_{\omega}$ and
$\mathcal{K}'$.$\square$

\begin{example}\label{e-b-ch}
By proposition \ref{hf-p-4} strong additivity of the
$\chi$-capacity holds for arbitrary channel $\Psi$ and the channel
$\Phi_{p}^{a}$ considered in the example in \cite{Sh-7} with
arbitrary probability density function $p(t)$ and $a\leq+\infty$.
This implies in particular that the classical capacity the channel
$\Phi_{p}^{a}$ with arbitrary constraint coincides with the
$\chi$-capacity.
\end{example}

\subsection{Representation for the CCoOE}

The convex closure of the output entropy (CCoOE) of a quantum
channel is an important characteristics related to the classical
capacity of this channel \cite{Sh-3}. This notion also plays
essential role in the theory of entanglement: an important
entanglement measure of a state of a composite quantum system --
the Entanglement of Formation (EoF) -- can be defined as the CCoOE
of a partial trace \cite{B&Co}.

By proposition \ref{prop-2} the CCoOE of a quantum channel
$\Phi\in\mathfrak{F}_{=1}(\mathcal{H},\mathcal{H}')$ is
represented by the expression
\begin{equation}\label{CCoOE-gen}
\overline{\mathrm{co}}H_{\Phi}(\rho)=\inf_{\mu\in\widehat{\mathcal{P}}_{\{\rho\}}}
\int_{\mathrm{extr}\mathfrak{S}(\mathcal{H})}H_{\Phi}(\sigma)
\mu(d\sigma),\quad \rho\in\mathfrak{S}(\mathcal{H}).
\end{equation}

In \cite{Sh-3} it is shown that for arbitrary state $\rho$ with
finite output entropy $H_{\Phi}(\rho)$ the infimum in this
expression can be taken only over atomic measures, which means
that
\begin{equation}\label{CCoOE-spec}
\overline{\mathrm{co}}H_{\Phi}(\rho)=\inf_{\{\pi_{i},\rho_{i}\}\in\widehat{\mathcal{P}}_{\{\rho\}}}
\sum_{i}\pi_{i}H_{\Phi}(\rho_{i}),
\end{equation}
(where the infimum is over all countable ensembles
$\{\pi_{i},\rho_{i}\}$ of pure states with the average state
$\rho$.)

But validity of expression (\ref{CCoOE-spec}) for arbitrary state
$\rho$ remains open question. The second example in remark 2 in
\cite{Sh-6} shows that positive answer on this question can not be
obtained by using only general analytical properties of the
(output) entropy. For given channel $\Phi$ validity of expression
(\ref{CCoOE-spec}) for arbitrary state $\rho$ is equivalent to
lower semicontinuity of the right side of this expression on the
input state space $\mathfrak{S}(\mathcal{H})$.

Thus in the case of  general quantum channel $\Phi$ it is
necessary to use representation (\ref{CCoOE-gen}), which involves
optimization over \textit{all} measures with given barycenter
$\rho$. This provides some technical problems in dealing with
CCoOE. Moreover this expression looks unnatural from the physical
point of view since for given state $\rho$ with \textit{finite}
mean energy, produced in a physical experiment, the above
optimization involves measures supported by states with
\textit{infinite} mean energy.\footnote{Any countable ensemble
having the average state with finite mean energy consists of
states with finite mean energy.}

In this subsection we obtain representation for the CCoOE of an
arbitrary quantum channel
$\Phi\in\mathfrak{F}_{=1}(\mathcal{H},\mathcal{H}')$ as a limit of
increasing sequence of \textit{continuous} bounded convex
functions on $\mathfrak{S}(\mathcal{H})$ defined via the
expressions similar to (\ref{CCoOE-spec}).

Let $n>1$ be fixed natural number. Consider the function
$$
H_{\Phi}^{n}(\rho)=
-\sum_{i=1}^{n}\lambda_{i}\log\lambda_{i}+\left(\sum_{i=1}^{n}\lambda_{i}\right)\log\left(\sum_{i=1}^{n}\lambda_{i}\right)=
H\left(\left\{\frac{\lambda_{i}}{\sum_{i=1}^{n}\lambda_{i}}\right\}_{i=1}^{n}\right),
$$
where $\{\lambda_{i}\}_{i=1}^{n}$ is the set of $n$ maximal
eigenvalues of the state $\Phi(\rho)$, which can be called
truncated output entropy. By lemma 4 in \cite{L-2} the sequence
$\{H_{\Phi}^{n}\}$ of continuous bounded functions on
$\mathfrak{S}(\mathcal{H})$ is nondecreasing and pointwise
converges to the output entropy $H_{\Phi}$.

Let 
$$
\check{H}^{n}_{\Phi}(\rho)=\inf_{\{\pi_{i},\rho_{i}\}\in\widehat{\mathcal{P}}_{\{\rho\}}}
\sum_{i}\pi_{i}H_{\Phi}^{n}(\rho_{i}).
$$

By proposition 5 in \cite{Sh-6} the function
$\check{H}^{n}_{\Phi}(=(H^{n}_{\Phi})_{*})$ is the convex
\textit{continuous} extension of the function
$\mathrm{extr}\mathfrak{S}(\mathcal{H})\ni\rho\mapsto
H^{n}_{\Phi}(\rho)$ to the set
$\mathfrak{S}(\mathcal{H})$.\footnote{Since in general case the
function $H^{n}_{\Phi}$ is not concave on
$\mathfrak{S}(\mathcal{H})$ we can not assert that
$\check{H}^{n}_{\Phi}=\overline{\mathrm{co}}H^{n}_{\Phi}$.}

The sequence $\{\check{H}^{n}_{\Phi}\}_{n}$ of convex continuous
bounded functions on $\mathfrak{S}(\mathcal{H})$ is increasing and
majorized by the function $\overline{\mathrm{co}}H_{\Phi}$. The
results of the previous section make possible to prove the
following observation.\footnote{It is nontrivial since the set
$\mathfrak{S}(\mathcal{H})$ is not compact.}

\begin{property}\label{CCoOE-representation}
\textit{For arbitrary channel
$\Phi\in\mathfrak{F}_{=1}(\mathcal{H},\mathcal{H}')$ and arbitrary
state $\rho\in\mathfrak{S}(\mathcal{H})$ the following relation
holds
$$
\overline{\mathrm{co}}H_{\Phi}(\rho)=\lim_{n\rightarrow+\infty}\check{H}^{n}_{\Phi}(\rho)=
\lim_{n\rightarrow+\infty}\inf_{\{\pi_{i},\rho_{i}\}\in\widehat{\mathcal{P}}_{\{\rho\}}}
\sum_{i}\pi_{i}H^{n}_{\Phi}(\rho_{i}).
$$}
\end{property}

\begin{remark}\label{on-EoF-representation}
This proposition does not imply validity of expression
(\ref{CCoOE-spec}). There exists an increasing sequence $\{f_{n}\}$
of concave continuous bounded functions on
$\mathfrak{S}(\mathcal{H})$ converging to the (concave lower
semicontinuous) bounded function $f$ such that
$$
\lim_{n\rightarrow+\infty}\inf_{\{\pi_{i},\rho_{i}\}\in\widehat{\mathcal{P}}_{\{\rho\}}}
\sum_{i}\pi_{i}f_{n}(\rho_{i})=0\quad\textup{and}\quad
\inf_{\{\pi_{i},\rho_{i}\}\in\widehat{\mathcal{P}}_{\{\rho\}}}
\sum_{i}\pi_{i}f(\rho_{i})=1
$$
for some state $\rho\in\mathfrak{S}(\mathcal{H})$ (see the second
example in remark 2 in \cite{Sh-6}).
\end{remark}

\textbf{Proof.} By the above observation it is sufficient to show
that
\begin{equation}\label{local-ineq}
\liminf_{n\rightarrow+\infty}\check{H}^{n}_{\Phi}(\rho)\geq
\overline{\mathrm{co}}H_{\Phi}(\rho)
\end{equation}
for arbitrary state $\rho\in\mathfrak{S}(\mathcal{H})$.

Let $\{P_{n}\}$ be a sequence of projectors in
$\mathfrak{B}(\mathcal{H}')$, increasing to the unit operator
$I_{\mathcal{H}}$, such that $\mathrm{rank}P_{n}=n$. Consider the
sequence $\{\Phi_{n}(\cdot)=P_{n}\Phi(\cdot)P_{n}\}$ of operations
in $\mathfrak{F}_{\leq 1}(\mathcal{H},\mathcal{H}')$.

Let $\rho$ be an arbitrary pure state in
$\mathfrak{S}(\mathcal{H})$. If $\{\lambda_{i}\}_{i=1}^{n}$ and
$\{\lambda^{n}_{i}\}_{i=1}^{n}$ are sets of maximal eigenvalues
(in decreasing order) of the operators $\Phi(\rho)$ and
$\Phi_{n}(\rho)$ then the Ritz principle implies
$\lambda_{i}\geq\lambda^{n}_{i}$ for each $i=\overline{1,n}$.
Hence by using (\ref{H-fun-eq}) we obtain
$$
H^{n}_{\Phi}(\rho)=H\left(\left\{\frac{\lambda_{i}}{\sum_{i=1}^{n}\lambda_{i}}\right\}_{i=1}^{n}\right)\geq
H(\{\lambda_{i}\}_{i=1}^{n})\geq
H(\{\lambda^{n}_{i}\}_{i=1}^{n})=H_{\Phi_{n}}(\rho).
$$
It follows that
$$
\inf_{\{\pi_{i},\rho_{i}\}\in\widehat{\mathcal{P}}_{\{\rho\}}}
\sum_{i}\pi_{i}H^{n}_{\Phi}(\rho_{i})\geq
\inf_{\{\pi_{i},\rho_{i}\}\in\widehat{\mathcal{P}}_{\{\rho\}}}
\sum_{i}\pi_{i}H_{\Phi_{n}}(\rho_{i}),\quad
\forall\rho\in\mathfrak{S}(\mathcal{H}).
$$
Since the function $H_{\Phi_{n}}$ is concave continuous and
bounded on $\mathfrak{S}(\mathcal{H})$ corollary 10 in \cite{Sh-6}
implies that the right side of the above inequality coincides with
$\overline{\mathrm{co}}H_{\Phi_{n}}(\rho)$.

The sequence $\{\Phi_{n}\}$ satisfies condition A in proposition
\ref{cont-cond}. Hence for arbitrary state $\rho$ in
$\mathfrak{S}(\mathcal{H})$ we obtain
$$
\liminf_{n\rightarrow+\infty}\inf_{\{\pi_{i},\rho_{i}\}\in\widehat{\mathcal{P}}_{\{\rho\}}}
\sum_{i}\pi_{i}H^{n}_{\Phi}(\rho_{i})\geq
\lim_{n\rightarrow+\infty}\overline{\mathrm{co}}H_{\Phi_{n}}(\rho)=\overline{\mathrm{co}}H_{\Phi}(\rho),
$$
which means (\ref{local-ineq}). $\square$

\begin{corollary}\label{CCoOE-u-c}
\textit{Let $\Phi\in\mathfrak{F}_{=1}(\mathcal{H},\mathcal{H}')$ be
an arbitrary channel and $\mathcal{A}$ be such compact subset of
$\mathfrak{S}(\mathcal{H})$ that the output entropy $H_{\Phi}$ is
continuous on $\mathcal{A}$. Then the increasing sequence
$\{\check{H}^{n}_{\Phi}\}$ of continuous functions converges to the
function $\overline{\mathrm{co}}H_{\Phi}$ uniformly on
$\mathcal{A}$.}
\end{corollary}

\textbf{Proof.} Proposition 7 in \cite{Sh-3} implies continuity of
the function $\overline{\mathrm{co}}H_{\Phi}$ on the set
$\mathcal{A}$. Hence the assertion of the corollary follows from
proposition \ref{CCoOE-representation} and Dini's lemma. $\square$

Corollary \ref{CCoOE-u-c} shows that for arbitrary Gaussian channel
$\Phi$ the sequence $\{\check{H}^{n}_{\Phi}\}$ provides uniform
approximation of the function $\overline{\mathrm{co}}H_{\Phi}$ on
the set of states with bounded mean energy (see the remark after
proposition 3 in \cite{H-Sh-2}).

Let $\mathcal{H}$ and $\mathcal{K}$ be separable Hilbert spaces.
Consider the channel
$\Theta:\mathfrak{S}(\mathcal{H}\otimes\mathcal{K})\ni\omega\mapsto\mathrm{Tr}_{\mathcal{K}}\omega\in\mathfrak{S}(\mathcal{H})$.
The Entanglement of Formation of a state
$\omega\in\mathfrak{S}(\mathcal{H}\otimes\mathcal{K})$ can be
defined by (cf.\cite{Sh-3})
$$
E_{\mathrm{F}}(\omega)=\overline{\mathrm{co}}H_{\Theta}(\omega)=\inf_{\mu\in\widehat{\mathcal{P}}_{\{\omega\}}}
\int_{\mathrm{extr}\mathfrak{S}(\mathcal{H}\otimes\mathcal{K})}H_{\Theta}(\sigma)\mu(d\sigma).
$$

Proposition \ref{CCoOE-representation} implies that
$$
E_{\mathrm{F}}(\omega)=\lim_{n\rightarrow+\infty}\check{H}^{n}_{\Theta}(\omega)=
\lim_{n\rightarrow+\infty}\inf_{\{\pi_{i},\omega_{i}\}\in\widehat{\mathcal{P}}_{\{\omega\}}}
\sum_{i}\pi_{i}H^{n}_{\Theta}(\omega_{i}).
$$
This proves the conjecture that $E_{\mathrm{F}}$ is a function of
class $\widehat{P}(\mathfrak{S}(\mathcal{H}\otimes\mathcal{K}))$
(cf.\cite{Sh-6}).

Corollary \ref{CCoOE-u-c} implies that the above convergence is
uniform on the set of states of composite system with bounded mean
energy.

\section{Appendix}

The following compactness criterion for subsets of
$\mathfrak{T}_{1}(\mathcal{H})$ can be proved by simple
modification of the arguments used in the proof of the compactness
criterion for subsets of $\mathfrak{S}(\mathcal{H})$, presented in
the Appendix in \cite{H-Sh-2}.

\textbf{Proposition.} \textit{The closed subset $\mathcal{A}$ of
$\mathfrak{T}_{1}(\mathcal{H})$ is (trace norm) compact if and only
if for arbitrary $\varepsilon>0$ there exists a finite rank
projector $P_{\varepsilon}$ such that
$\mathrm{Tr}(I_{\mathcal{H}}-P_{\varepsilon})A<\varepsilon$ for all
$A\in\mathcal{A}$.}

\textbf{Corollary.} \textit{Let $\mathcal{A}$ and $\mathcal{B}$ be
subsets of $\mathfrak{T}_{1}(\mathcal{H})$ and
$\mathfrak{T}_{1}(\mathcal{K})$ correspondingly. The subset
$\mathcal{A}\otimes \mathcal{B}$ of
$\mathfrak{T}_{1}(\mathcal{H}\otimes\mathcal{K})$ consisting of
all operators $C$ such that
$\mathrm{Tr}_{\mathcal{K}}C\in\mathcal{A}$ and
$\mathrm{Tr}_{\mathcal{H}}C\in\mathcal{B}$ is compact if and only
if the sets $\mathcal{A}$ and $\mathcal{B}$ are compact.}

\textbf{Proof.} Compactness of the set $\mathcal{A}\otimes
\mathcal{B}$ implies compactness of the sets $\mathcal{A}$ and
$\mathcal{B}$ due to continuity of partial trace.

Let $\mathcal{A}$ and $\mathcal{B}$ be compact. By the above
proposition for arbitrary $\varepsilon>0$ there exist finite rank
projectors $P_{\varepsilon}$ and $Q_{\varepsilon}$ such that
$$
\mathrm{Tr}P_{\varepsilon}A>\mathrm{Tr}A-\varepsilon,\;\forall
A\in\mathcal{A},\quad \mathrm{and}\quad
\mathrm{Tr}Q_{\varepsilon}B>\mathrm{Tr}B-\varepsilon,\;\forall
B\in\mathcal{B}.
$$
Since $C^{\mathcal{H}}=\mathrm{Tr}_{\mathcal{K}}C\in\mathcal{A}$
and $C^{\mathcal{K}}=\mathrm{Tr}_{\mathcal{H}}C\in\mathcal{B}$ for
arbitrary $C\in\mathcal{A}\otimes \mathcal{B}$ we have
$$
\begin{array}{c}
 \mathrm{Tr}((P_{\varepsilon}\otimes Q_{\varepsilon})\cdot C)=\mathrm{Tr}((P_{\varepsilon}\otimes
I_{\mathcal{K}})\cdot C)-\mathrm{Tr}(P_{\varepsilon}\otimes
(I_{\mathcal{K}}-Q_{\varepsilon}))\cdot C)\\\\\geq
\mathrm{Tr}P_{\varepsilon}C^{\mathcal{H}}-
\mathrm{Tr}(I_{\mathcal{K}}-Q_{\varepsilon})C^{\mathcal{K}}>\mathrm{Tr}C-2\varepsilon.
\end{array}
$$
The above proposition implies compactness of the set
$\mathcal{A}\otimes \mathcal{B}$.$\square$

\end{document}